\documentclass[a4paper]{jpconf}
\usepackage{graphicx}

\usepackage[dvipsnames]{xcolor}

\usepackage{amsmath,bm}
\usepackage{dsfont}
\usepackage{subcaption}

\usepackage[switch, modulo]{lineno}

\begin{document}
\title{Beam test performance of a highly granular silicon tungsten calorimeter technical prototype for the ILD}

\author{F~Jiménez~Morales, on behalf of the CALICE Collaboration}

\address{Laboratoire Leprince-Ringuet, CNRS, École polytechnique, Institut Polytechnique de Paris, 91120 Palaiseau, France.}

\ead{fabricio.jimenez@llr.in2p3.fr}

\begin{abstract}
A highly granular silicon-tungsten electromagnetic calorimeter (SiW-ECAL) is the reference design of the ECAL for International Large Detector concept, one of the two detector concepts for the future International Linear Collider.
Prototypes for this type of detector are developed within the CALICE Collaboration.
The technological prototype addresses technical challenges such as integrated front-end electronics or compact layer and readout design.
A stack of 7 layers was compiled and tested at DESY test beam facilities in 2017. We present preliminary results on the properties of the electromagnetic showers.
An outline on the next steps is given.
Finally, we illustrate the first steps of the digitization concept on simulations of the prototype.
\end{abstract}


\section{Introduction}\label{sec:introduction}

A central demand of the next generation of high-energy particle colliders, such as the International Linear Collider (ILC), is achieving a highly precise reconstruction of the final states produced in the collisions.
The ILC is proposed to deliver collisions of polarized $e^{+}e^{-}$ beams with center-of-momentum energies ranging from 250 GeV to 1 TeV; if realized, the ILC will become the most powerful linear collider and lepton collider ever built, where Higgs bosons will be produced copiously in a clean environment.
For pursuing the physics program of the ILC, two multi-purpose detector concepts are being developed: the International Large Detector (ILD) and the Silicon Detector (SiD)~\cite{Behnke:2013lya}.
The guiding principle that underlies these detector concepts is the Particle Flow (PF) Algorithm~\cite{Brient:2002gh,Morgunov:2004ed}, where single-particle separation and precise energy measurement are achieved by combining calorimetric and tracking information.
Calorimeters that are compact and highly granular are then a requirement for implementing PF techniques; such devices are currently under development within the CALICE collaboration ~\cite{Sefkow:2015hna}.


This work revolves around a silicon-tungsten (SiW) electromagnetic calorimeter (ECAL) prototype for ILD.
The SiW-ECAL prototype is part of the CALICE research and development (R\&D) program to construct an ECAL that complies with the ILD baseline design specifications.
The ECAL for the ILD is a sampling calorimeter that uses silicon as the active material and tungsten as the absorber material, reaching 24 radiation lengths (24 $X_0$) in the barrel region in 23 centimeters for 26-30 active layers.
This allows for the construction of an ECAL that is compact and highly granular, with small silicon sensor cells (squares of 5$\times$5 mm$^2$), totaling around 100 million channels for the ILD.

The ECAL for the ILD will profit from the special collision setup at the ILC, where $e^+e^-$ bunch crossings happen around every 200 ms: the electronics are active during an acquisition window of $\sim$1-2 ms, and shut down the rest ($\sim$99\%) of the time.
This technique, that reduces overall power consumption is known as power pulsing. 


We show preliminary results obtained from data recorded at a test beam session from 2017 and the current status of the implementation of digitization effects in simulations that are being prepared for comparison studies.

\section{The SiW-ECAL prototype}\label{sec:proto}

Within the R\&D program for the SiW-ECAL, we distinguish two main versions of the prototypes: the physics prototype, and the technological prototype.
The physics prototype was the first to be developed and provided a proof-of-concept for PF calorimetry~\cite{Adloff:2011ha,Anduze:2008hq,Adloff:2008aa,Adloff:2010xj,CALICE:2011aa,Bilki:2014uep}, where the electronic components were placed outside the interaction region, used an external trigger system, and had no constraints on the power consumption.
After this successful first iteration, the following years have been devoted to the technological prototype to address engineering challenges at ILD, such as electronic components inside the detector, implementing power pulsing, achieving required compactness, among others.

In this work, we present preliminary results using data recorded with the technological prototype in 2017.
At that time, the prototype consisted of a stack of seven layers, each of which contained an Active Sensor Unit (ASU), an interface card to the data acquisition system, the tungsten absorber, carbon supports and shielding.
Each ASU, of surface area $18\times18$ cm$^2$, is composed of 16 readout chips, each of which has 64  cells (for a total of 1024 per ASU, and 7168 in the whole seven-layer prototype) of size $5.5\times5.5$ mm$^2$; further, four silicon wafers of 320 µm thickness are glued onto the ASU.
This represents an improvement in the silicon lateral granularity and thickness with respect to the physics prototype, where the cell square surface measured $10\times10$ mm$^2$, and were 500 µm thick.
The PCB version used for the 2017 prototype ASUs is the FEV11, that comprises SKIROC2~\cite{Callier:2011zz} chips for the readout.
An adjustable trigger threshold value is set in the SKIROC2, which is needed as the ILC will not provide a central trigger; if the energy deposited by an incident particle in a cell produces a signal that passes the threshold value, all 64 cell energy values are recorded by the chip as cell hits.


\section{Electromagnetic showers using 2017 test beam data}

The physics program for the test beam session of 2017 at DESY comprised several items to be studied.
Before the test beam session, a dedicated commissioning procedure was developed for the prototype, where cell thresholds and masks were adjusted and set.
At the test beam facility, calibration runs without tungsten absorber plates using a beam of 3 GeV positrons were performed, first perpendicular to the prototype, and then at an angle of 45 degrees (the latter using only six layers), see references~\cite{Irles:2018uum,Kawagoe:2019dzh}.
Different positron beam energies and tungsten configurations in the layers were used to study the response of the prototype, where electromagnetic showers develop; in this section we report preliminary results on this item.


The beam used was a positron beam perpendicular to the layers of the fully-equipped prototype, with energies of 1, 2, 3, 4, 5, and 5.8 GeV.
The absorber material was arranged in three configurations, for capturing different parts of the shower development:
\begin{itemize}
    \item Configuration 1: 0.6, 1.2, 1.8, 2.4, 3.6, 4.8 and 6.6 $X_0$\,,
    \item Configuration 2 1.2, 1.8, 2.4, 3.6, 4.8, 6.6 and 8.4 $X_0$\,,
    \item Configuration 3 1.8, 2.4, 3.6, 4.8, 6.6, 8.4 and 10.2 $X_0$\,.
\end{itemize}

We select the reconstructed events (in a time window of 200 ns) and their hits with two criteria, based on the number of layers hits in the events and the hit energies in all events.
The event selection consists in requiring the number of layers hits to be greater than five, as actual shower events would lead to record, in principle, hits in all layers. 
A gaussian distribution is fit in a window near the noise peak in the energy spectrum, i.e. near zero for values calibrated to the Minimum Ionizing Particle (MIP) value; then, only hits whose energy is greater than six standard deviations above the mean of the gaussian distribution are selected.
By imposing these requirements we filter out noise and are left with a data set that can be used to model the development of electromagnetic showers that develop in the detector.




The shower energy modeling is carried out in two separate components: the transversal part (T), that is used per layer, and the longitudinal part (L, see reference~\cite{1975NucIM.128..283L}):
\begin{equation}\label{eq:longtrans}
    \text{T: } A (f G(\bm\mu, \mathds{1}\sigma_1) + (1-f) G(\bm\mu, \mathds{1}\sigma_2))\,, \hspace{2em} \text{L: } \frac{\textnormal d E}{\textnormal d t} = E_0 b \frac{(bt)^{a-1} e^{-bt}}{\Gamma(a)}\,.
\end{equation}
For the transversal part, we use on each layer a double two-dimensional gaussian distribution with shared mean $\bm\mu=(\mu_x, \mu_y)$ for each layer, where $A$ is a global constant, $f$ is the relative strength of the core gaussian ($0<f<1$), and the diagonal entries of the correlation matrices that accommodate the narrow shower core ($\sigma_1$) with a peripheral halo ($\sigma_2$).
In the longitudinal model, $t$ is the absorber depth, and $a, b$ parameters that depend on the material (W) and the energy scale of the interaction.
(See Section 33.5 of reference~\cite{Tanabashi:2018oca} for more detail.)



\begin{figure}[ht]
    \begin{subfigure}{0.5\textwidth}
        \includegraphics[width=0.95\textwidth]{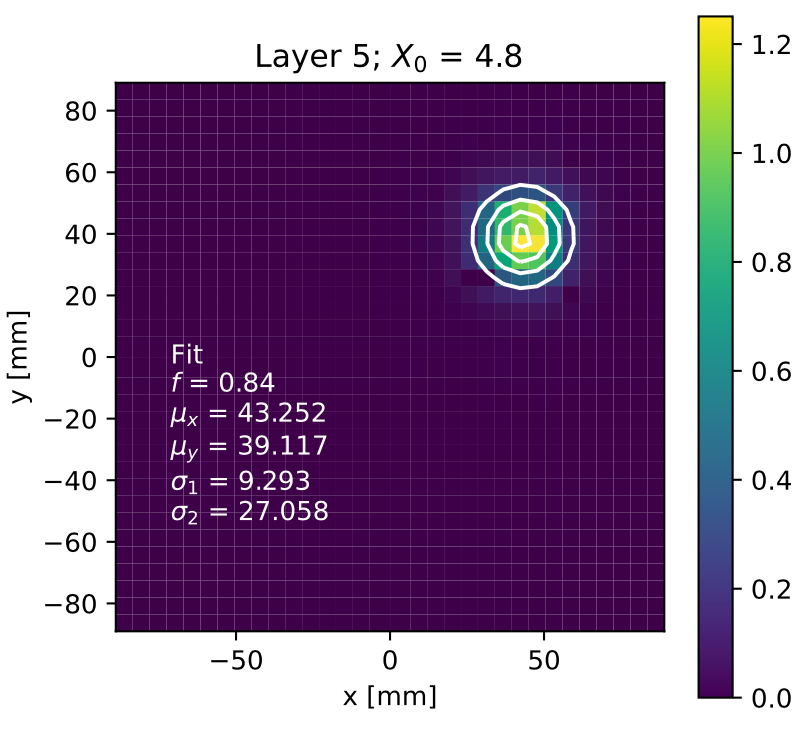}
    \end{subfigure}
    \begin{subfigure}{0.5\textwidth}
        \includegraphics[width=0.95\textwidth]{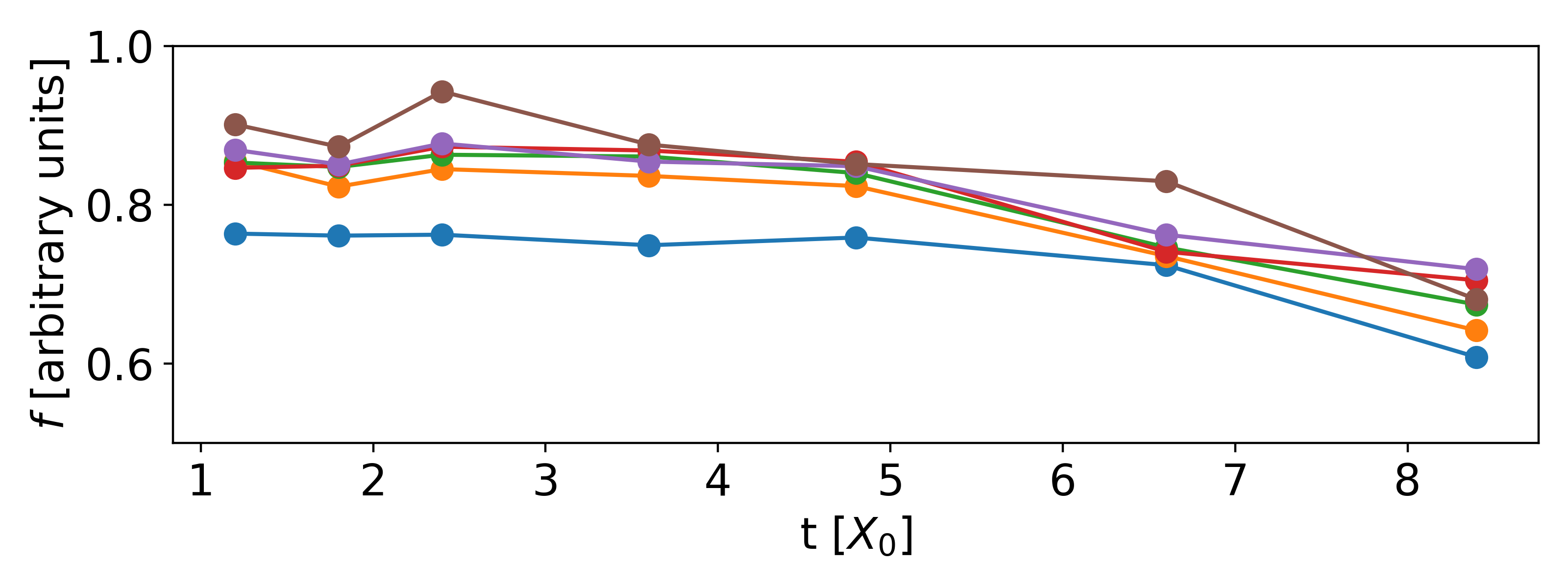}\\
        \includegraphics[width=0.95\textwidth]{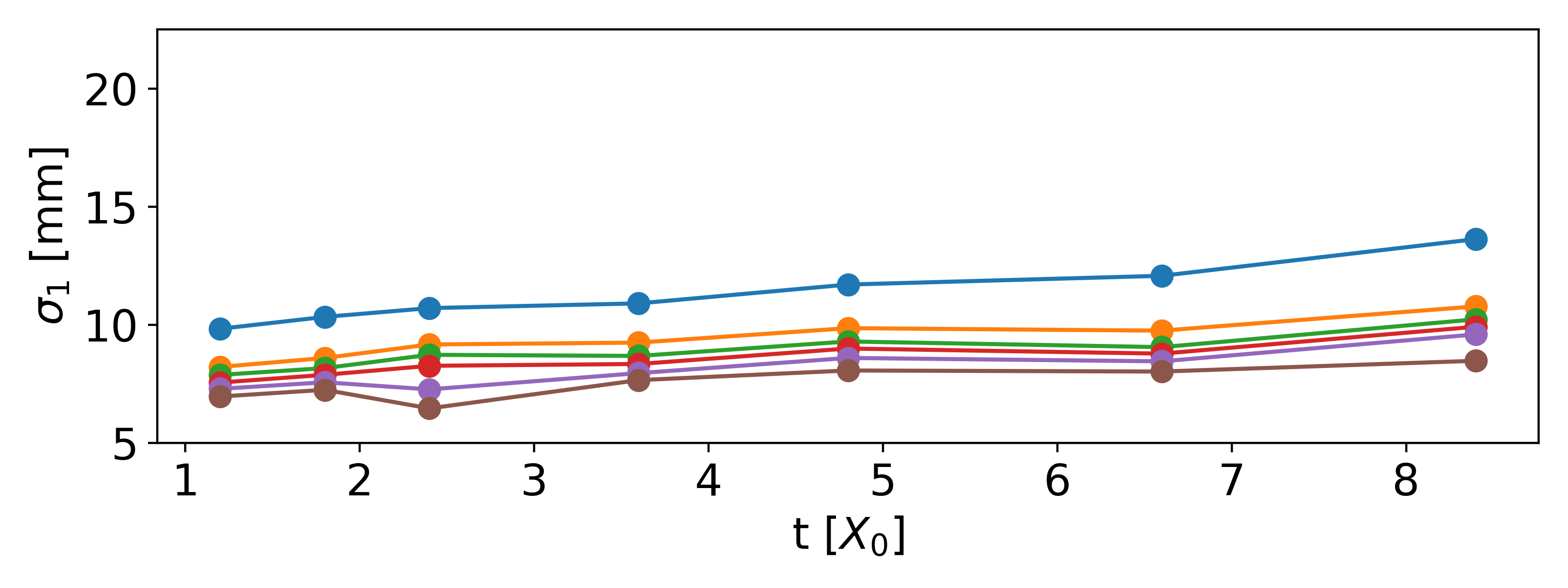}
    \end{subfigure}
    \vspace{0.1em}
    \begin{subfigure}{0.5\textwidth}
        \includegraphics[width=0.95\textwidth]{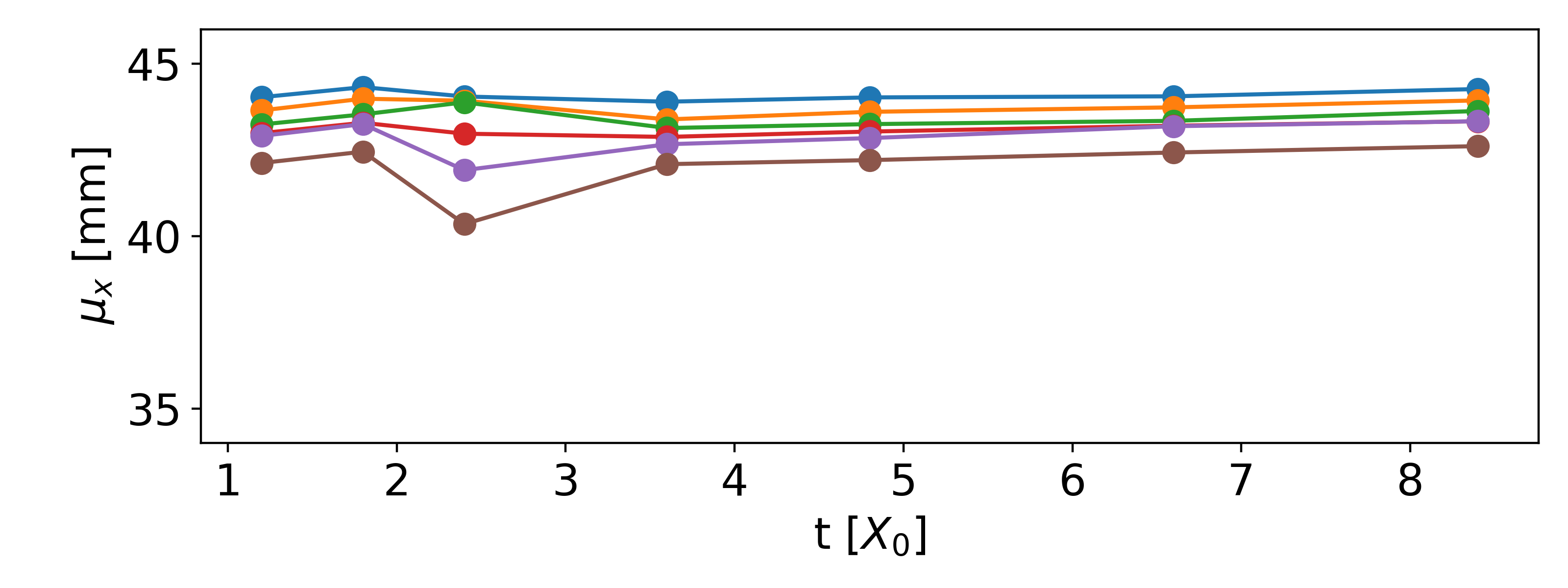}\\
        \includegraphics[width=0.95\textwidth]{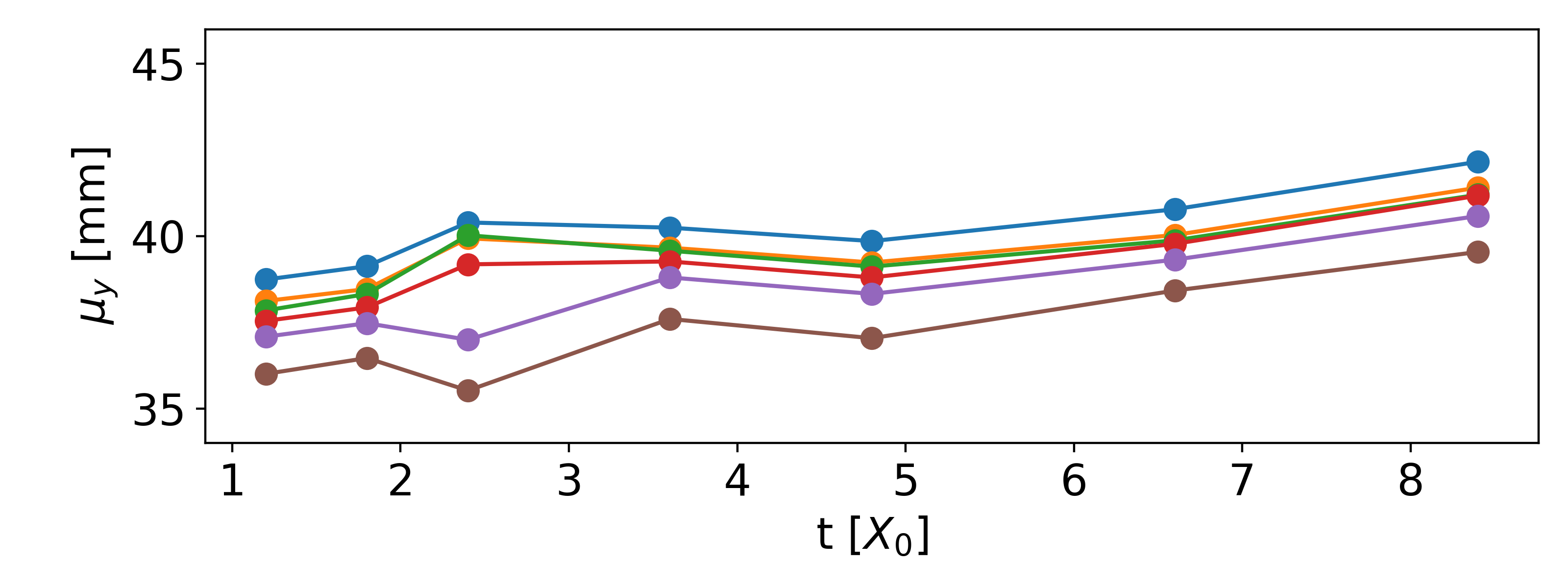}
    \end{subfigure}
    \begin{subfigure}{0.5\textwidth}
        \includegraphics[width=0.95\textwidth]{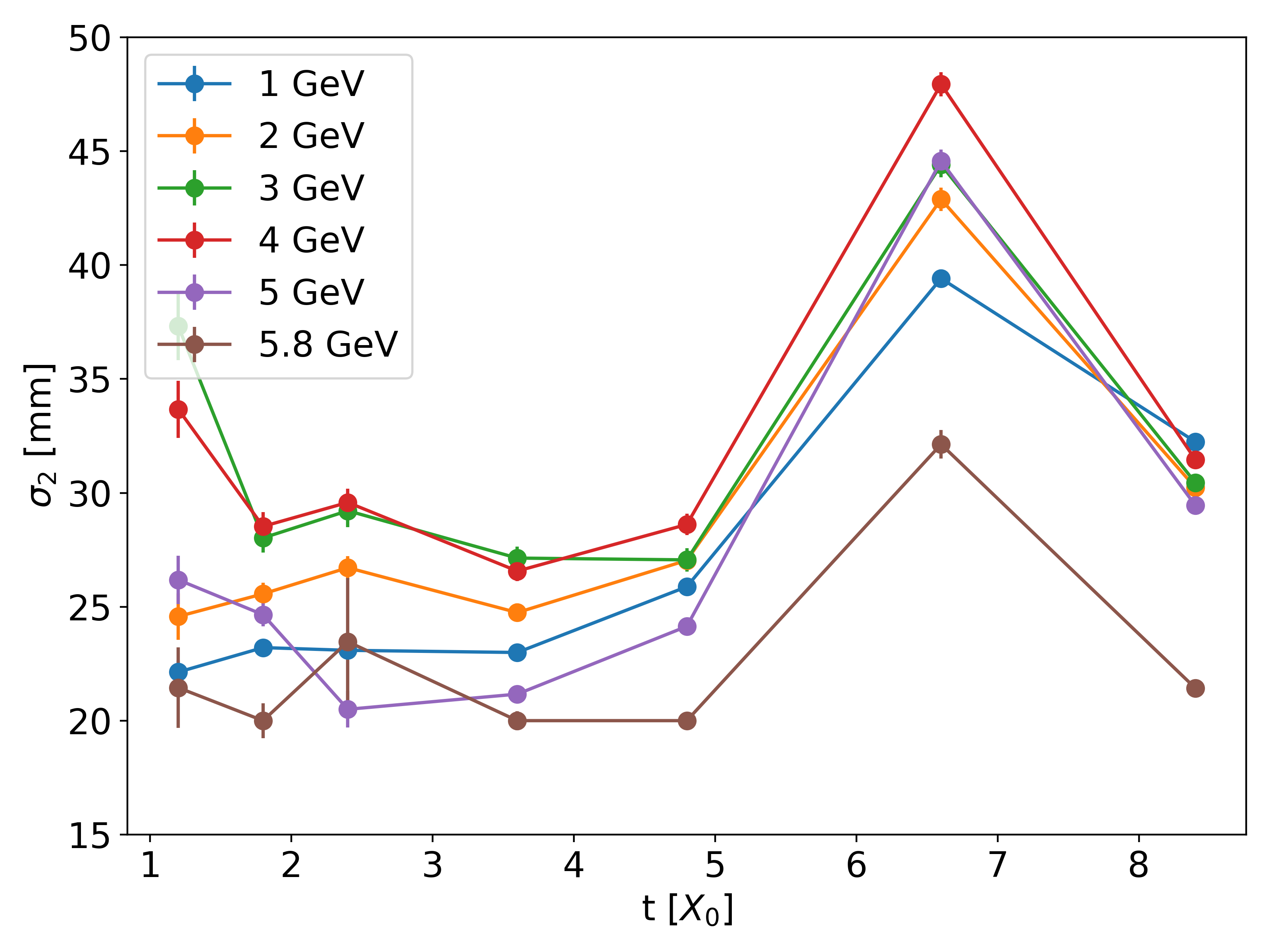}
    \end{subfigure}
    \caption{Upper left: average energy deposits (color scale in MIPs) on the 5th layer in configuration 2 with a beam of 3 GeV, with its double gaussian model fit (solid white lines). The rest of the plots show the evolution of the model parameters with W depth for configuration 2 (all energies). On the center and bottom left, the components of the model mean ($\mu_x, \mu_y$) are shown. On the right, the signal fraction $f$, and widths $\sigma_1$ and $\sigma_2$ are shown from top to bottom.}
    \label{fig:long_parameters}
\end{figure}

\begin{figure}[ht]
    \begin{subfigure}{0.5\textwidth}
        \includegraphics[width=0.95\textwidth]{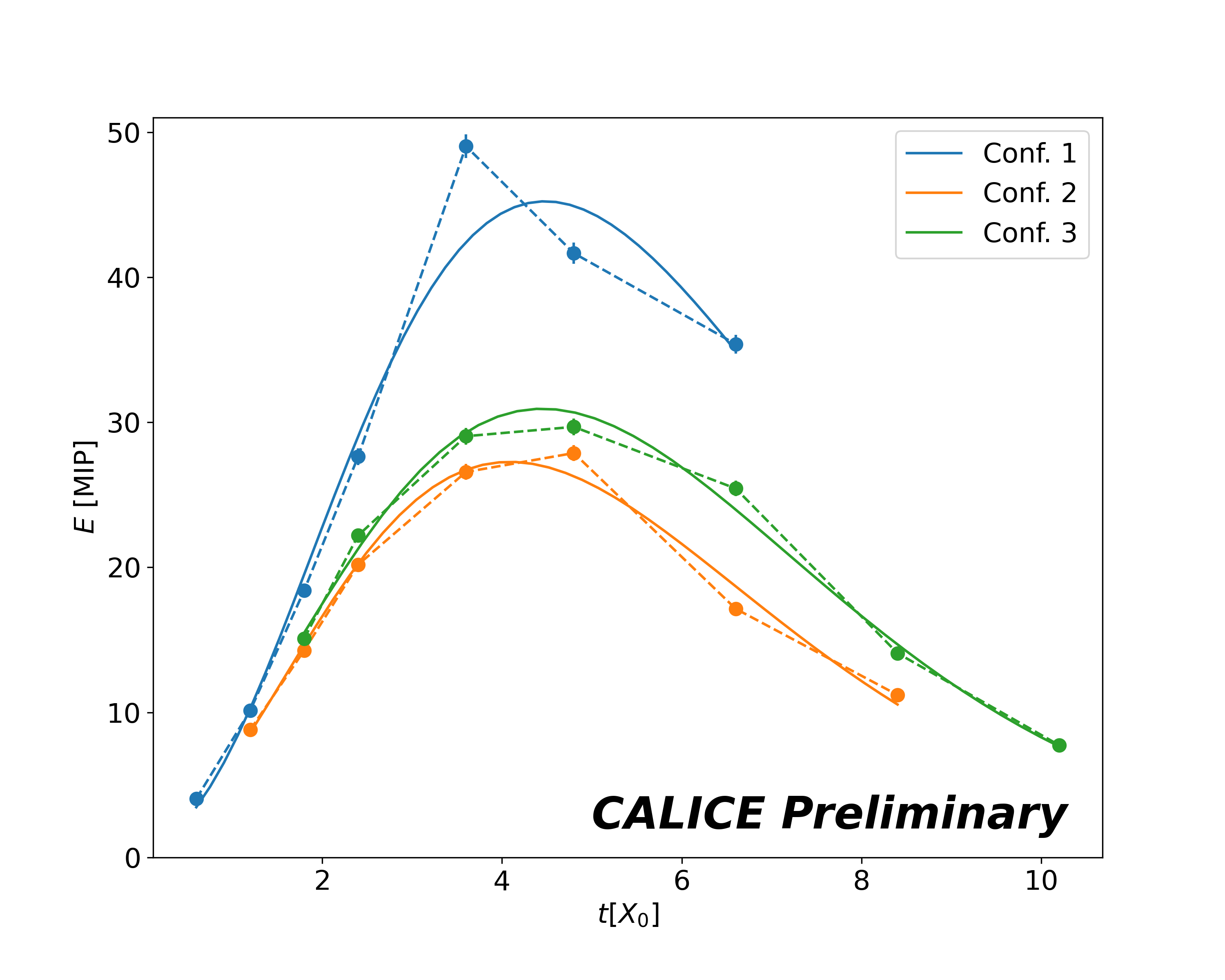}
    \end{subfigure}
    \begin{subfigure}{0.5\textwidth}
        \includegraphics[width=0.95\textwidth]{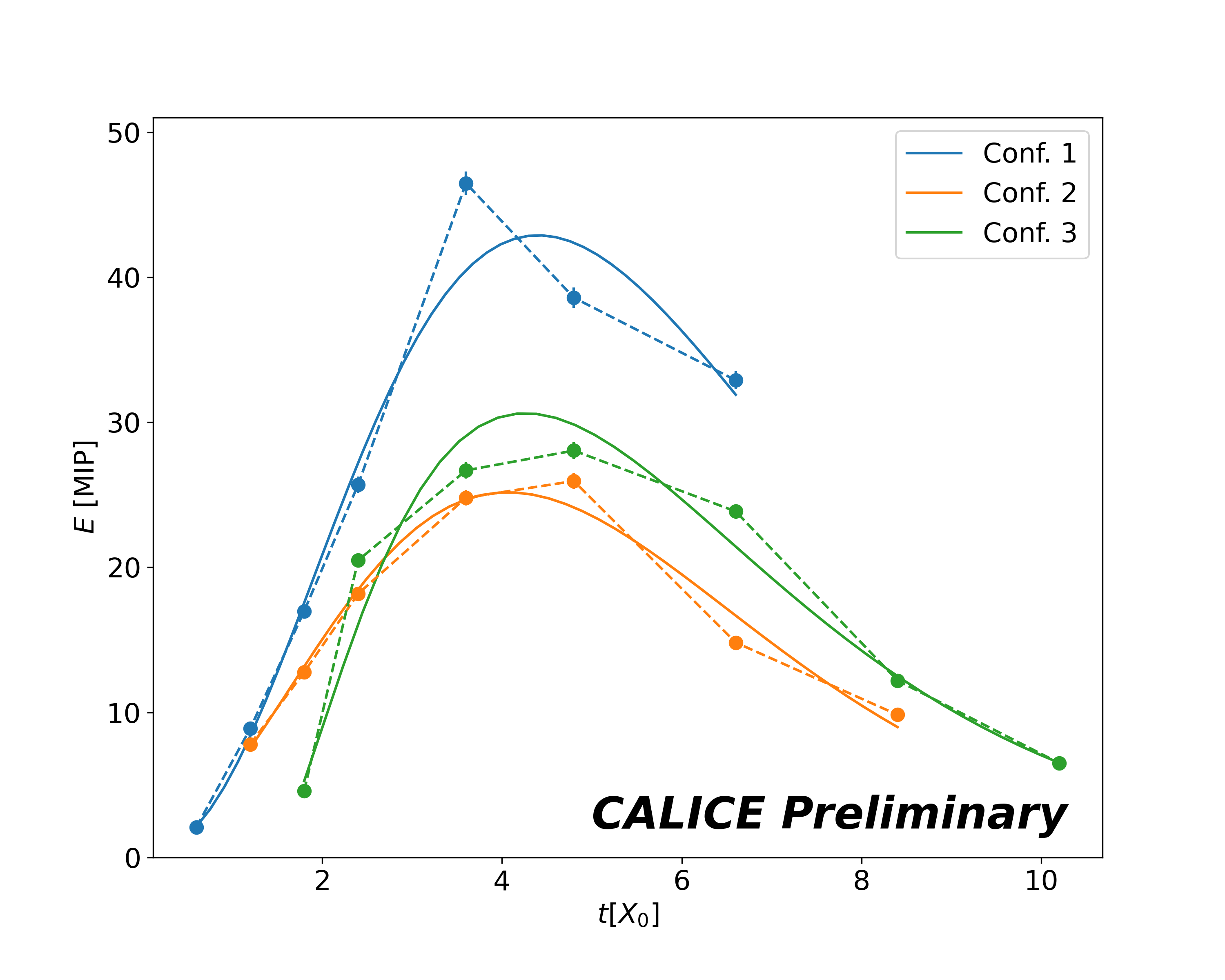}
    \end{subfigure}
    \caption{Left: average shower energy per layer (dots) as a function of tungsten depth $t$, with its longitudinal profile fit (solid lines, eq. L~(\ref{eq:longtrans})). Right: integral from double gaussian model (dots) and respective fit (solid lines). Using data for 3 GeV positrons, all configurations.}
    \label{fig:long_energies}
\end{figure}

In figure~\ref{fig:long_parameters} an example of the average hit energy in one layer is shown, with a respective double gaussian model fit, as well as the evolution of the double gaussian model parameters in the different layers.
An advantage of having a fitted model at hand is that it allows for mitigating the effect of data imperfections, e.g. due to dead or masked cells; thus the model can be used as a prompt for the average energy distribution in the layer.
We see several effects in the evolution of the parameters: the increase in the width of the core ($\sigma_1$) as the shower develops in the detector, an overall constant value of double gaussian horizontal mean ($\mu_x$) and an increasing trend in the vertical mean ($\mu_y$) which hints that the prototype layers were not perfectly perpendicular to the beam in that direction; the excess in $\sigma_2$ for the next-to-last layer could be due to a reflection with the last layer of absorber, but needs to be investigated in future work.
The longitudinal profile is fit on two quantities: the average energy deposited by all hits on each layer and on the integral of the double gaussian model fit, as it is shown in the plots on figure~\ref{fig:long_energies}.

This exploratory work and first fits in test beam data need further developments in several aspects.
The description of electromagnetic showers can be improved by e.g. defining a shower axis and having a radial description for the transversal component or further using a full 3D model instead of separating the longitudinal and transversal components.
Also, validation studies on the robustness of the results with respect to the selection used should be performed.
Finally, the current analysis should also be extended to study individual shower events, and include energy response functions over the cell surface.

During 2019-20 a new iteration of the prototype with up to 22 layers with a dimension of $\sim18\times18\times25$ cm$^3$ was compiled and is ready for a test beam session.


\section{Digitization for the SiW-ECAL}\label{sec:digitization}

A complete study of the prototype demands a comparison of the data collected with simulations, that broadly comprises two steps.
First, we have the simulation of the physics of the incident beam and its interaction with the prototype layers.
Second, the digitization, which consists in including the effects of the readout electronic components.
A comparison between the simulated data obtained at that stage with those collected by the prototype is possible, where an assessment of the detector performance and potential improvements to the simulation can be obtained, which is essential for the advancement of the prototype and final detector design.

Simulations of the 2017 SiW-ECAL prototype at the test beam have been prepared using the DD4hep framework~\cite{frank_markus_2021_4923786}, where the interaction of the incoming beam with all the prototype components is modeled, following the same setup as the test beam described above as well as using incident beams of electrons, positrons (3 GeV) and muons (0.4, 4 and 40 GeV) without tungsten absorbers.

\begin{figure}
    \begin{subfigure}{0.5\textwidth}
        \includegraphics[width=0.9\textwidth]{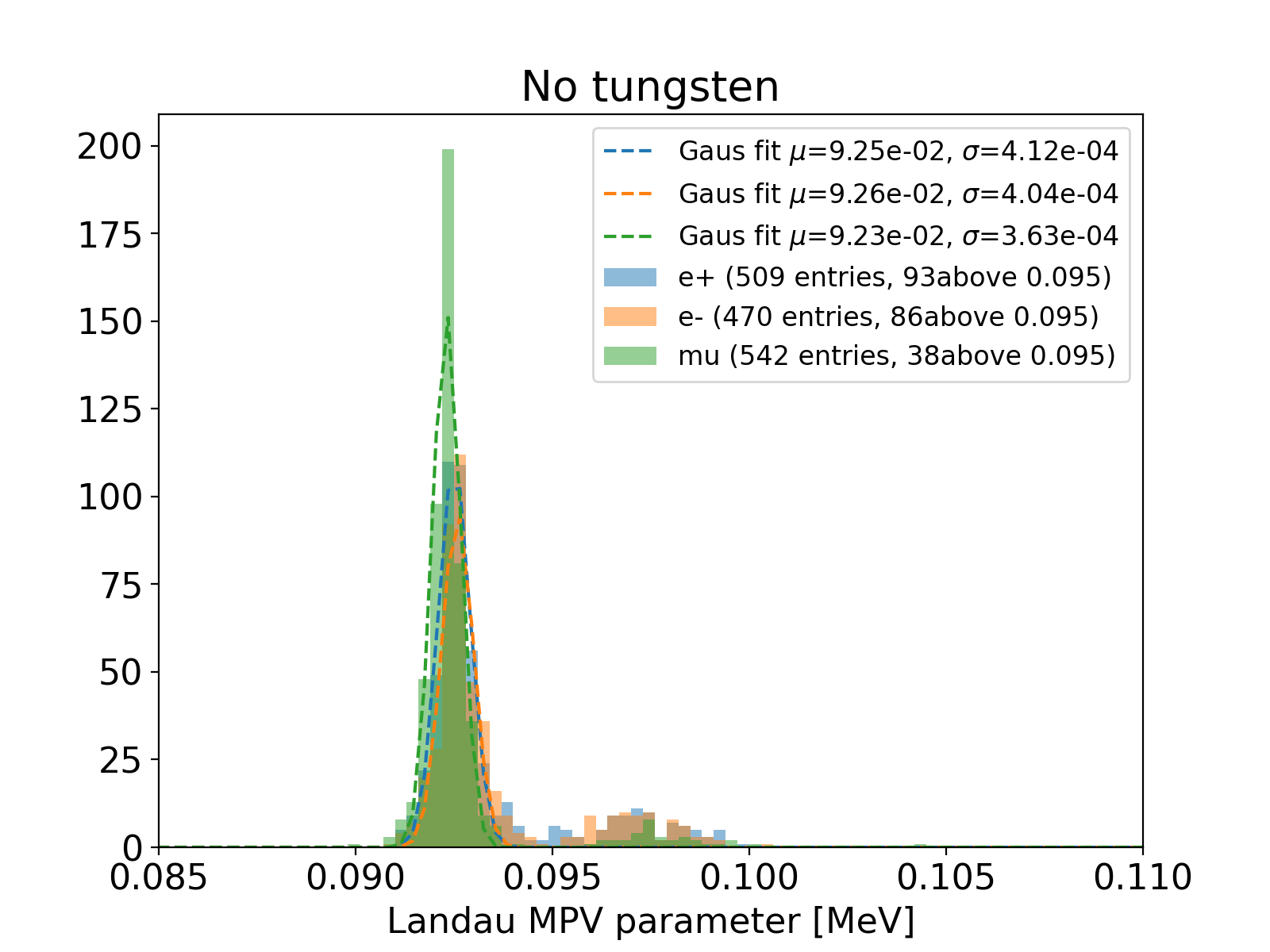}
    \end{subfigure}
    \begin{subfigure}{0.5\textwidth}
        \includegraphics[width=0.9\textwidth]{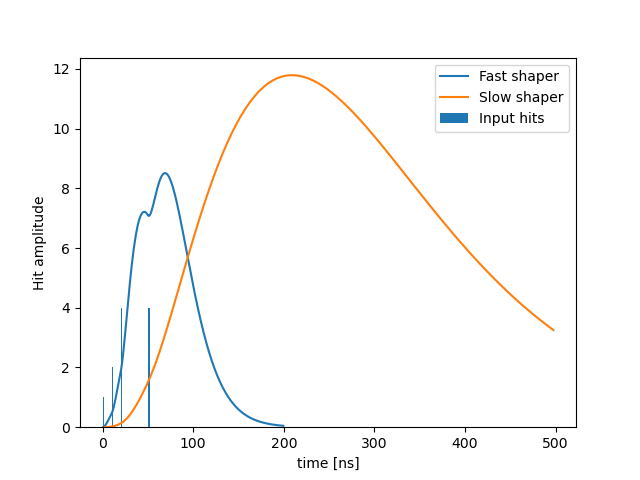}
    \end{subfigure}
    \caption{Left: histogram of Landau MPV values from the cell energy response (bars) for $e^-, e^+,$ and $\mu^-$, with gaussian fits (dashed lines). Right: illustration of input hits (blue bars, in arbitrary units) and the response Slow and Fast Shaper filters (blue and orange lines).}
    \label{fig:mpvfits}
\end{figure}

The first step of the digitization is to convert the value of the energy deposit into the MIP scale.
For that, we use as a reference the simulated average response to a muon beam at 40 GeV crossing normally the prototype layers, without tungsten.
Each cell that contained at least 1000 hits (for each 10000 events) was used to build an energy spectrum, that was modeled using a Landau function.
A histogram of the values of the landau function most probable value (MPV) obtained from all cells was fit with a gaussian distribution, with a mean at 9.23$\times10^{-5}$ GeV with a width of 0.04$\times10^{-5}$ GeV; this preliminary mean value is taken as the MIP value and used for conversion.
The same procedure was followed for the case of 3 GeV beams of electrons and positrons in the absence of tungsten, for reference.
The distributions and their fits are shown in figure~\ref{fig:mpvfits}, where a number (quoted in the figure) of MPV values are sensibly above the bulk of the distribution, that are thought to be an artifact of the setup used to build the cell energy histogram, and should be investigated in detail.

As a second step of the digitization, we model the information path followed by the signal pulse from the cell though the SKIROC2 chip.
The incoming pulse is first pre-amplified and then fed into the fast and slow shapers, which are represented by second order CR-RC filters, with typical times ($\tau$) equal to 30 and 180 ns respectively.
When the fast shaped signal rises above the threshold set in the chip, it triggers the recording of the amplitude of the slow shaped signals in analog memory, after a programmable delay, ideally to sample them at their maximum.
Figure~\ref{fig:mpvfits} shows an illustration of the response of the shapers to input hits.
Further effects need to be included in the digitization, such as adding a noise term in the cells and readout, the Analog-to-Digital signal Conversion, and adding time smearing.
Future studies should investigate the impact of the digitization steps with respect to the input simulation and with the data recorded.



\section*{Acknowledgements}

This project has received funding from the European Union’s Horizon 2020 Research and Innovation programme under Grant Agreement no. 654168.
The measurements leading to these results have been performed at the Test Beam Facility at DESY Hamburg (Germany), a member of the Helmholtz Association (HGF).

\section*{References}
\bibliographystyle{plain}
\bibliography{mybib}

\begin{thebibliography}{10}

\bibitem{Behnke:2013lya}
Halina Abramowicz et~al.
\newblock {The International Linear Collider Technical Design Report - Volume
  4: Detectors}.
\newblock 6 2013.

\bibitem{Adloff:2011ha}
C.~Adloff, J.~Blaha, J.~J. Blaising, C.~Drancourt, A.~Espargiliere, R.~Galione,
  N.~Geffroy, Y.~Karyotakis, J.~Prast, and G.~Vouters.
\newblock {Tests of a particle flow algorithm with CALICE test beam data}.
\newblock {\em JINST}, 6:P07005, 2011.

\bibitem{Adloff:2008aa}
C.~Adloff et~al.
\newblock {Response of the CALICE Si-W electromagnetic calorimeter physics
  prototype to electrons}.
\newblock {\em Nucl. Instrum. Meth. A}, 608:372--383, 2009.

\bibitem{Adloff:2010xj}
C.~Adloff et~al.
\newblock {Study of the interactions of pions in the CALICE silicon-tungsten
  calorimeter prototype}.
\newblock {\em JINST}, 5:P05007, 2010.

\bibitem{CALICE:2011aa}
C.~Adloff et~al.
\newblock {Effects of high-energy particle showers on the embedded front-end
  electronics of an electromagnetic calorimeter for a future lepton collider}.
\newblock {\em Nucl. Instrum. Meth. A}, 654:97--109, 2011.

\bibitem{Bilki:2014uep}
B.~Bilki et~al.
\newblock {Testing hadronic interaction models using a highly granular
  silicon\textendash{}tungsten calorimeter}.
\newblock {\em Nucl. Instrum. Meth. A}, 794:240--254, 2015.

\bibitem{Brient:2002gh}
Jean-Claude Brient and Henri Videau.
\newblock {The Calorimetry at the future e+ e- linear collider}.
\newblock {\em eConf}, C010630:E3047, 2001.

\bibitem{Callier:2011zz}
S.~Callier, F.~Dulucq, C.~de~La~Taille, G.~Martin-Chassard, and
  N.~Seguin-Moreau.
\newblock {SKIROC2, front end chip designed to readout the Electromagnetic
  CALorimeter at the ILC}.
\newblock {\em JINST}, 6:C12040, 2011.

\bibitem{Anduze:2008hq}
The~CALICE collaboration.
\newblock Design and electronics commissioning of the physics prototype of a
  si-w electromagnetic calorimeter for the international linear collider.
\newblock {\em Journal of Instrumentation}, 3(08):P08001--P08001, aug 2008.

\bibitem{frank_markus_2021_4923786}
Markus Frank, Frank Gaede, Marko Petric, and Andre Sailer.
\newblock Aidasoft/dd4hep: v01-17-00, June 2021.
\newblock webpage: http://dd4hep.cern.ch/.

\bibitem{Irles:2018uum}
Adri\'an Irles.
\newblock {Latest R\&D news and beam test performance of the highly granular
  SiW-ECAL technological prototype for the ILC}.
\newblock {\em JINST}, 13(02):C02038, 2018.

\bibitem{Kawagoe:2019dzh}
K.~Kawagoe et~al.
\newblock {Beam test performance of the highly granular SiW-ECAL technological
  prototype for the ILC}.
\newblock {\em Nucl. Instrum. Meth. A}, 950:162969, 2020.

\bibitem{1975NucIM.128..283L}
Egidio {Longo} and Ignazio {Sestili}.
\newblock {Monte Carlo calculation of photon-initiated electromagnetic showers
  in lead glass}.
\newblock {\em Nuclear Instruments and Methods}, 128(2):283--307, October 1975.

\bibitem{Morgunov:2004ed}
V.~{Morgunov} and A.~{Raspereza}.
\newblock {Novel 3D Clustering Algorithm and Two Particle Separation with Tile
  HCAL}.
\newblock {\em arXiv e-prints}, page physics/0412108, December 2004.

\bibitem{Sefkow:2015hna}
Felix Sefkow, Andy White, Kiyotomo Kawagoe, Roman P\"oschl, and Jos\'e Repond.
\newblock {Experimental Tests of Particle Flow Calorimetry}.
\newblock {\em Rev. Mod. Phys.}, 88:015003, 2016.

\bibitem{Tanabashi:2018oca}
M.~Tanabashi et~al.
\newblock {Review of Particle Physics}.
\newblock {\em Phys. Rev. D}, 98(3):030001, 2018.

\end{thebibliography}


\end{document}